\documentclass[11pt,twoside]{article}
\usepackage{asp2010}

\resetcounters

\begin{document}

\title{DrizzlePac 2.0 - Introducing New Features}
\author{Roberto~J.~Avila$^1$, Warren~Hack$^1$, Mihai~Cara$^1$, David~Borncamp$^1$, Jennifer~Mack$^1$, Linda~Smith$^{1,2}$, and Leonardo~Ubeda$^1$
\affil{$^1$Space Telescope Science Institute,\\
     3700 San Martin Drive, Baltimore, MD 21218, USA}
\affil{$^2$European Space Agency,\\
     3700 San Martin Drive, Baltimore, MD 21218, USA}}

\begin{abstract}
The DrizzlePac package includes tasks for aligning and drizzling images taken with the Hubble Space Telescope. We present this release which includes new features that facilitate image alignment, sky matching, and adds support for new time dependent distortion solutions of the ACS instrument. The TweakReg task now includes capabilities for automatically aligning images which form part of a mosaic. In addition, new parameters make it easier to reject cosmic rays and other spurious detections from source catalogs used for alignment. The Astrodrizzle task has been improved with a new sky matching algorithm which makes producing mosaics easier than ever before. This new version supports an improved version of the ACS/WFC time-dependent distortion correction. There are also improvements to the GUI interfaces and some behind the scene bug fixes.
\end{abstract}

\section{Introduction}
DrizzlePac\footnote{\url{http://drizzlepac.stsci.edu}} is an STScI software package that contains a suite of tasks for aligning and drizzling images taken with the Hubble Space Telescope \citep{gonzaga_2012, hack_2012}. In the last year, the DrizzlePac development team has been working on improvements based on feedback from the community. We concentrated on two major areas for improvements; usability improvements and adding support for large datasets. In the following sections we describe the concerns in each of these areas that required attention and how we updated the software to address them.

\section{Usability Improvements}
Several enhancements were identified that would make the software more usable. We list these enhancements in order of priority below:

\subsection{Alignment}
The TweakReg task was created as a wrapper around the process of refining the world coordinate system (WCS) alignment in image headers. The task finds sources in the input images, aligns the images in WCS space, and updates the headers. In practice, this process is difficult to implement when the ratio of cosmic rays to real sources is high. When the input and reference images come from different filters or cameras, the image properties will be different and currently TweakReg uses a single set of source-finding parameters for all images. Additionaly, image artifacts like CCD bleeding, wings of bright stars, persistence in IR images, and edge effects generate spurious detections in the source catalog. To get around these problems, prior versions of the software required time consuming work-arounds and iterative solutions in the workflow (eg. \citealp{avila_2014}). To address these problems we implemented the following changes to TweakReg:

\begin{itemize}
\item Separate sets of source-finding parameters for input and reference images.
\item New parameters in the source-finding step to weed out extended sources and hot pixels. With this feature there is less of a need to use CR-cleaned images to make source catalogs.
\item Support for the more DS9 region shapes and the capability to declare both inclusion and exclusion regions. This helps to minimize spurious detections caused by image artifacts. 
\end{itemize}

\subsection{Sky subtraction}
Measuring and matching the sky level is important to the drizzle process. When the background has not been properly matched between frames, the cosmic-ray rejection may be compromised. In the most severe cases, overly aggressive flagging can impact the photometric accuracy of the drizzled products. Additionally, when building mosaics, discontinuities between tiles appear when the background has not been properly matched. This is especially a problem when a bright, extended object dominates one of the tiles. 

Figure \ref{sky_comp} shows an example of mis-matched sky in a 3$\times$3 mosaic. The left half of the plot shows the results produced by the current sky measuring algorithm and the right half shows the new version. A cut across the lower third of the mosaic is over-plotted on the images. The left plot shows the discontinuity between the two bottom left tiles. It also shows how the sky in each bottom tile has been mis-calculated because the local minimum is used as the sky value, without any regard to the information provided by the other images in the mosaic.

The sky measuring step in AstroDrizzle has been re-written. The code calculates the sky using the overlap region between a pair of images. Then the sky background is computed that produces the best match between the sky values of the various overlaps between all pairs of images, in the least squares sense. 

\articlefigure[scale=1.0]{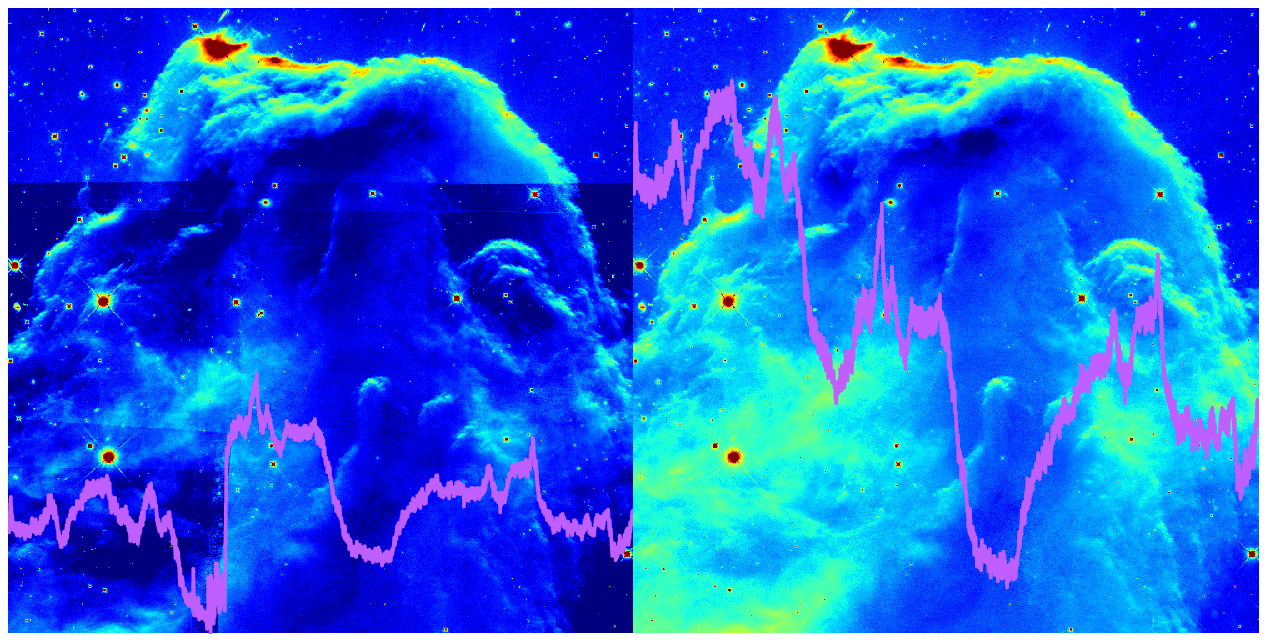}{sky_comp}{Mosaics drizzled using the current (left) and new (right) sky matching algorithms. The mis-matched skies are clearly visible on the left. Overplotted on the two images, using identical axes, is a 50 pixels wide cut across the bottom third of the mosaics.}

\subsection{Automation}
Because the DrizzlePac software is written in Python, the tasks are scriptable so that pipelines can be built around it. Even so, there were several issues that prevented the software from running in a fully automated mode. 

\begin{itemize}
\item TweakReg diagnostic plots can now be saved for later viewing so that the user no longer needs to run the task interactively.
\item STWCS can now apply refined WCS solutions contained within headerlet files in batch mode.
\item Fixed clash between parameter names in TweakReg and Imagefind.
\end{itemize}

\section{Large Dataset Support}
We also identified improvements that would benefit projects with large sets of images. These can be projects that involve mosaics or deeps fields with single pointings. Each of these modes presents different challenges which we addressed in the following ways:

\subsection{Mosaic building}
Building mosaics can be a difficult endeavor. First of all, to properly align tiles, one needs to iterate through a series of steps made up of aligning images, drizzling images to create a reference source catalog, then aligning the next tile (e.g. \citealt{mack_2013}). To solve this problem, TweakReg now automatically adds the sources from an input image that has been aligned to the reference catalog and continues with the next image on the list. With this capability, images in a mosaic can be aligned with one call to TweakReg. TweakReg can also identify the order in which images should be aligned based on the size of the overlap region between images.   

When building large mosaics, the single drizzle step makes large, mostly empty, images which are stored on the hard drive or in memory. To address this the single drizzle step now implements an image compression algorithm that reduces the footprint these images have on the system resources.

\subsection{Deep fields}
Projects that observe deep fields may encounter problems when trying to align images taken years apart. The ACS team is actively working on a new solution for the time dependency component of the distortion. Additionally, the team is planning on implementing an algorithm that removes skew from the astrometric residuals. This new version of the software supports an improved ``beta" version of the ACS/WFC time-dependent distortion correction \citep{borncamp_2014}, with further improvements coming in the future.

\section{Bug fixes and future improvements}
In addition to these feature enhancements, several bug fixes were fixed in the TEAL GUI. 

Currently AstroDrizzle spends $\sim$50\% of its computing time in only 2 of the 8 processing steps (blot and final drizzle). In the future, we hope to introduce parallel processing capabilities to these steps.

\section{Obtaining the new software}
This new version of the software is not yet part of UREKA and is only available under the SSBX testing environment. To obtain the new software, please go to the Science Software Branch website of the Space Telescope Science Institute. Detailed installation instructions can be found there.

\url{http://ssb.stsci.edu/ssb_software.shtml}

\bibliography{P2-18}

\end{document}